\documentclass[prb,twocolumn,aps,superscriptaddress,showpacs,floatfix]{revtex4}
\usepackage{amsmath}
\usepackage{bm}
\usepackage{amsfonts}
\usepackage{graphicx}
\usepackage{color}
\usepackage{tikz}
\begin{document}
\title{Tunable Majorana fermion from Landau quantization in 2D topological
superconductors}

\author{R.S. Akzyanov}
\affiliation{Moscow Institute of Physics and Technology, Dolgoprudny, Moscow Region, 141700 Russia}
\affiliation{Institute for Theoretical and Applied Electrodynamics, Russian
Academy of Sciences, Moscow, 125412 Russia}
\affiliation{Dukhov Research Institute of Automatics, Moscow, 127055 Russia }

\author{A.L. Rakhmanov}
\affiliation{Moscow Institute of Physics and Technology, Dolgoprudny, Moscow Region, 141700 Russia}
\affiliation{Institute for Theoretical and Applied Electrodynamics, Russian
Academy of Sciences, Moscow, 125412 Russia}
\affiliation{Dukhov Research Institute of Automatics, Moscow, 127055 Russia }

\affiliation{CEMS, RIKEN, Saitama, 351-0198, Japan}

\author{A.V. Rozhkov}
\affiliation{Moscow Institute of Physics and Technology, Dolgoprudny, Moscow Region, 141700 Russia}
\affiliation{Institute for Theoretical and Applied Electrodynamics, Russian
Academy of Sciences, Moscow, 125412 Russia}
\affiliation{CEMS, RIKEN, Saitama, 351-0198, Japan}

\author{Franco Nori}
\affiliation{CEMS, RIKEN, Saitama, 351-0198, Japan}
\affiliation{Department of Physics, University of Michigan, Ann Arbor, MI 48109-1040, USA}

\begin{abstract}
We study Majorana fermions in a two-dimensional topological superconductor
placed in a transverse magnetic field $B$. We consider a topological
insulator/superconductor heterostructure and a two-dimensional $p$-wave
superconductor. A single field-generated vortex creates two Majorana
fermions, one of which is hosted at the vortex core. The wave function of
the second Majorana state is localized in the superconductor volume along a
circle of radius
$r^*\propto B^{-1}$
centered at the vortex core. In the case of many vortices, the sensitivity of
$r^*$
to the magnetic field $B$ may be used to control the coupling between the
Majorana fermions. The latter property could be an asset for quantum
computations.
\end{abstract}
\date{\today}

\maketitle
\section{Introduction}
Majorana fermions (MFs) in condensed matter became the focus of many
studies~\cite{Alicea,Leijnse,Beenakker,Franz,sato_arxiv},
especially in connection with the future possibility of
topologically-protected
computation~\cite{kitaev_braid,tqc}.
A necessary operation for such a computation is braiding, which could be
performed~\cite{Akhmerov_anyon_braid,Clarke_braid,Beenakker_braid}
by tuning the pairwise interaction between the MFs (``non-topological"
qubit
operations~\cite{tqc}
were also discussed in the literature,
Refs.~\onlinecite{bravyi,bonderson2010,flensberg2011,peng_zhang,
schmidt2013,kovalev2014}).
Different structures have been proposed as possible hosts for
MFs~\cite{stanecu_nanowire,fu_kane_device,feigel1,das_sarma_FI,tiwari1,AL,
me_edge,me,DasSarma1,you_nori,sato2009,sato2010}.
Preliminary experimental hints of Majorana states were observed in a
nanowire with strong spin-orbit coupling on top of a
superconductor~\cite{exp1},
in a ferromagnetic atoms
chain~\cite{atomic_chains},
and, most recently, in a topological insulator/superconductor (TI/SC)
heterostructure~\cite{hetero_exp_majorana}.

It is generally assumed that MFs localize at some heterogeneity separating
media with different topological numbers, which may be a vortex core in a
superconductor or superfluid, sample boundaries, interfaces in
heterostructures, etc. (see, e.g.,
Refs.~\onlinecite{Alicea,stone_edge}).
The attachment of the MF to such a physical ``defect" may be
disadvantageous for a number of reasons: the presence of physical
inhomogeneities in the system requires additional efforts at the
fabrication stage, it introduces extra disorder, and creates interfaces,
whose properties are difficult to control. In particular, binding a MF to
some point in space limits its control and manipulation options.

However, it is not a general law that MF must be localized near some
``topological defect". Below we consider two models of a two-dimensional
(2D) topological superconductor hosting a vortex. Systems of this type have
been studied
experimentally~\cite{williams_jj,snelder},
and hints of a Majorana state at the vortex core were
reported~\cite{hetero_exp_majorana}.
From the theory standpoint, these models are particularly interesting,
because, as it will be demonstrated below, they may serve as a platform
where a confinement mechanism unrelated to a physical heterogeneity is
realized. The MFs can arise only in pairs, since only a superposition of
two MFs has a physical
sense~\cite{Leijnse}.
Therefore, in addition to the well-studied MF at the
core~\cite{Volovik_vortex,ivanov,Kraus,AL},
a second, exterior, Majorana state can
emerge~\cite{me_edge}.
The second MF is not necessarily pinned by some interface or sample edge.
We show that it can be localized by a finite magnetic field. For a uniform
field $B$, the wave function of the exterior MF can be calculated exactly.
Its weight is centered at some field-dependent mesoscopic radius
$r^*\propto B^{-1}$,
away from the vortex center. In other words, the wave function is localized
at a circle of radius
$r^*$.
Varying the magnetic field, we may control
$r^*$.
This feature could be useful for the manipulation of Majorana
states. We will analyze the tunability of the pairwise splitting between
MFs in the case of two vortices.

The paper is organized as follows. In
Sec.~\ref{sec::ti-sc}
we discuss the generation of the Majorana fermions in the topological
insulator-superconductor heterostructure. In
Sec.~\ref{sec::pwave}
similar ideas are applied to $p$-wave superconductor. The discussion and
conclusions are in
Sec.~\ref{sec::disco}
and
Sec.~\ref{sec::summary},
respectively.

\begin{figure}[t!]
\center
\includegraphics[width=7 cm, height=5 cm]{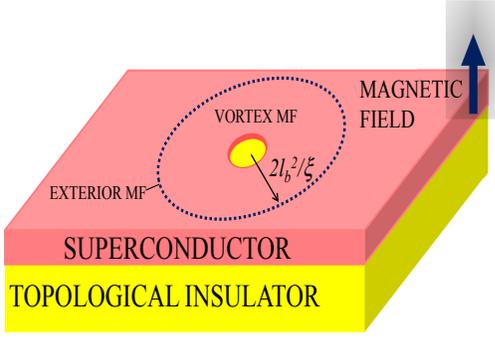}
\caption{(Color online) Topological insulator/superconductor
heterostructure with a vortex. A layer of topological insulator (yellow) is
covered by a superconducting film (pink) with
$s$-wave
order parameter. The yellow hole in the superconductor represents the
vortex with core-localized Majorana fermion. Dashed line corresponds to the
exterior Majorana fermion.
\label{shem}
}
\end{figure}

\section{TI/SC heterostructure}
\label{sec::ti-sc}

Given the recent
success~\cite{hetero_exp_majorana}
in the fabrication of TI/SC heterostructures, let us discuss this system
first. Our model describes the 2D states at the surface of the TI.
Proximity to the superconductor induces superconducting correlations in
these states. An external magnetic field $B$ inserts a vortex with integer
vorticity $l$ into the system, which is schematically shown in
Fig.~\ref{shem}.
The corresponding
Hamiltonian~\cite{fu_kane_device,AL,DasSarma1,me}
is a
4$\times$4 matrix
\begin{eqnarray}
\label{ham_ti_sc}
H
\!=\!
(v_{\rm F} \bm{\sigma}\! \cdot \!\textbf{p}\!-\!U)\tau_z\!
-\!
\frac{e v_{\rm F}}{c}(\bm{\sigma}\! \cdot\! \textbf{A})\tau_0\!
+
\!|\Delta(r)|\tau_x e^{il\phi\tau_z},
\end{eqnarray}
acting in a space of bi-spinor wave functions
\begin{eqnarray}
\psi = (u_\uparrow, u_\downarrow, v_\downarrow, - v_\downarrow)^T.
\end{eqnarray}
Above,
$\textbf{p}$
is the momentum operator,
$\bm{\sigma}$
is the vector of the three Pauli matrices acting in spin space,
$\tau_j$
are the Pauli matrices acting in the charge space, $U$ is the shift of the
Fermi level from the Dirac point, and
$v_{\rm F}$
is the Fermi velocity in the TI surface. We introduce polar coordinates
$(r,\phi)$
with the origin at the vortex core. In these coordinates, the
proximity-induced superconducting order parameter can be expressed as
\begin{eqnarray}
\label{order_parameter}
\Delta(\mathbf{r}) = |\Delta| f(r) e^{-i l \phi}.
\end{eqnarray}
The phase of
$\Delta(\mathbf{r})$
changes by
$2\pi l$
around the origin. The absolute value
$|\Delta({\bf r})| = |\Delta| f(r)$
varies with $r$: the function
$f(r)$
vanishes if
$r \ll \xi_{\rm SC}$,
and approaches unity when
$r \gg \xi_{\rm SC}$,
where
$\xi_{\rm SC}$
is the coherence length inside the superconductor. We assume that the
magnetic field is homogeneous (this is possible when the superconducting
layer is sufficiently thin). Below, the vector potential is chosen as
\begin{eqnarray}
A_\phi = -\frac{Br}{2},
\quad
A_r = A_z = 0.
\end{eqnarray}
The Hamiltonian $H$ possesses particle-hole symmetry: using the complex
conjugation operator $K$ one defines the particle-hole conjugation operator
\begin{eqnarray}
\Xi = \sigma_y \tau_y K,
\quad
\text{such that}
\quad
\Xi H\Xi = -H.
\end{eqnarray}
Thus, for any eigenfunction
$\psi_\varepsilon$,
satisfying
$H \psi_\varepsilon = \varepsilon \psi_\varepsilon$,
there is another eigenfunction
$\psi_{-\varepsilon} = \Xi \psi_\varepsilon$
with eigenenergy
$-\varepsilon$.
Since the MF wave function satisfies
\begin{eqnarray}
\Xi \psi_{\rm MF} = \psi_{\rm MF},
\end{eqnarray}
the corresponding eigenenergy is zero,
$\varepsilon_{\rm MF} = 0$.

We introduce a spinor
$F^\mu (r)=(f^{\mu}_1,f^{\mu}_2,f^{\mu}_3,-f^{\mu}_4)^T$
as
\begin{equation}
\label{F}
\psi =  \exp[-i\phi(l\tau_z-\sigma_z)/2+i\mu \phi]F^{\mu}(r),
\end{equation}
where the index
$\mu$
represents the total angular momentum of the state. In the case of a vortex
with single vorticity
($l = 1$),
the transformation~(\ref{F}) is well-defined only when
$\mu$
is an integer. We can express the equation
$H \psi = \varepsilon \psi$
as~\cite{me_edge}
\begin{eqnarray}
\label{final}
\nonumber
i \!\left (\frac d{dr}\!+\!\frac {2\mu+l+1}{2r}\!-\!\frac{r}{2 l_{\rm B}^2}\right )\!f^{\mu}_2\!+\!\frac{\Delta}{\hbar v_{\rm F}} f^{\mu}_3\!=\! \left(\frac{\varepsilon\!+\!U}{\hbar v_{\rm F}}\right)f^{\mu}_1,\quad
\\ \nonumber
i \!\left (\frac d{dr}\!-\!\frac {2\mu+l-1}{2r}\!+\!
\frac{r}{2 l_{\rm B}^2}\right)\!f^{\mu}_1\!-\!
\frac{\Delta}{\hbar v_{\rm F}} f^{\mu}_4\!=\!
\left(\frac{\varepsilon\!+\!U}{\hbar v_{\rm F}}\right)f^{\mu}_2,\quad
\\
i \!\left (\frac d{dr}\!+\!\frac {2\mu-l+1}{2r}\!+\!
\frac{r}{2 l_{\rm B}^2}\right)\!f^{\mu}_4\!+\!
\frac{\Delta}{\hbar v_{\rm F}} f^{\mu}_1\!=\!
\left(\frac{\varepsilon\!-\!U}{\hbar v_{\rm F}}\right)f^{\mu}_3,\quad
\\
\nonumber
i \!\left (\frac d{dr}\!-\!\frac {2\mu-l-1}{2r}\!-\!
\frac{r}{2 l_{\rm B}^2}\right)\!f^{\mu}_3\!-\!
\frac{\Delta}{\hbar v_{\rm F}} f^{\mu}_2\!=\!
\left(\frac{\varepsilon\!-\!U}{\hbar v_{\rm F}} \right)f^{\mu}_4,\quad
\end{eqnarray}
where
$l_{\rm B}^2 = \hbar c/ e B$
is the magnetic length. Majorana states exist only
when~\cite{me_edge,me}
$\varepsilon = 0$
and
$\mu=0$.
If $\varepsilon$ and $\mu$ vanish, we introduce the following linear
combinations
\begin{eqnarray}\label{Center}
 \nonumber
 X_1=if^{0}_1+f^{0}_4,\quad X_2=if^{0}_1-f^{0}_4,  \\
 Y_1=if^0_2+f^{0}_3,\quad Y_2=if^{0}_2-f^{0}_3,
\end{eqnarray}
for which
Eqs.~\eqref{final}
decouple into two independent systems of equations for
$(X_1,Y_2)$
and
$(X_2,Y_1)$.
Each of these systems can be reduced to the following equation
\begin{eqnarray}
\label{chi_eq}
\chi''+\frac{\chi'}{\bar r}
+
\chi\!\!
\left(
	{\bar U}^2
	-
	\frac{1}{\bar r^2}
	-
	\frac{\bar r^2}{4}
\right)
=0,
\end{eqnarray}
where the dimensionless variables
\begin{eqnarray}
{\bar r} = \frac{r}{l_{\rm B}},
\quad
{\bar U} = \frac{U}{\hbar \omega_{\rm c}}
\end{eqnarray}
are used, and the quantity
\begin{eqnarray}
\omega_{\rm c} = \frac{v_{\rm F}}{l_{\rm B}}
\end{eqnarray}
may be viewed as the cyclotron frequency for a massless relativistic
particle. Function $\chi$ is connected to
$Y_{1,2}$
as follows
\begin{eqnarray}
\label{y1}
Y_1=\chi ({\bar r})
\exp\left[ - \alpha \int_0^{\bar r} f({\bar r}')d {\bar r}'\right],
\\
\label{y2}
Y_2=\chi ({\bar r})
\exp\left[  \alpha \int_0^{\bar r} f({\bar r}')d {\bar r}'\right],
\\
\text{where}
\quad
\alpha=\frac{l_{\rm B}}{\xi},
\quad
\text{and}
\quad
\xi = \frac{\hbar v_{\rm F}}{|\Delta|}.
\end{eqnarray}
Length scale $\xi$ is the familiar coherence length due to the proximity
effect. The substitution
\begin{eqnarray}
\chi ({\bar r}) = \frac{g({\bar r}^2/2)}{\bar r}
\end{eqnarray}
transforms
Eq.~\eqref{chi_eq}
into a 1D Schr{\"o}dinger equation describing a quantum particle in an
attractive Coulomb potential. Using this analogy, it it easy to check that
Eq.~(\ref{chi_eq})
has a normalizable solution, provided that
\begin{eqnarray}
\label{N_int}
{\bar U}^2 = 2N \geq 0,
\quad
N\text{ is any non-negative integer,}
\end{eqnarray}
When these conditions hold, we solve
Eq.~(\ref{chi_eq})
and using
Eqs.~\eqref{F}~\eqref{final},
and~\eqref{Center}
we obtain the solution for the MF localized near the vortex core in the
form
\begin{eqnarray}
\label{vortex_m_1}
\psi_{\rm v}=B_{\rm v}
\exp\left[
		-\frac{i\pi}{4}
		-\frac{r^2}{4l_{\rm B}^2}
		-
		\int \limits_0^r f(r') \frac{dr'}{\xi}
\right]
\Psi_{\rm MF},
\\
\label{spinor_def}
\Psi_{\rm MF}(r, \phi)
=
\begin{bmatrix}
 L_N^{(0)}( {\bar r}^2/2 ) \\
-i ({\bar r}/{\bar U}) L_{N-1}^{(1)}( {\bar r}^2/2 ) e^{i\phi} \\
({\bar r}/{\bar U}) L_{N-1}^{(1)}({\bar r}^2/2 ) e^{-i\phi} \\
i L_N^{(0)}( {\bar r}^2/2) \\
\end{bmatrix},
\end{eqnarray}
where
$L_m^{(n)} (y)$
are generalized Laguerre
polynomials~\cite{NIST},
and the normalization coefficient
$B_{\rm v}$
is real. Note, if
\begin{eqnarray}
{\bar U}=N=0,
\quad
\text{then}
\quad
L_{-1}^{(1)}/{\bar U} \equiv 0.
\end{eqnarray}
To prove that
$\psi_{\rm v}$
is a MF, it is enough to check that
$\Xi \psi_{\rm v}=\psi_{\rm v}$.

The solution
$\psi_{\rm v}$
of
Eq.~(\ref{vortex_m_1})
corresponds to
Eq.~(\ref{y1}).
The solution that corresponds to
Eq.~(\ref{y2})
describes yet another MF state
\begin{eqnarray}
\label{vortex_m_2}
\psi_{\rm e}=B_{\rm e}
\exp\left[
		\frac{i\pi}{4}
		-\frac{r^2}{4l_{\rm B}^2}
		+
		\int \limits_0^r f(r') \frac{dr'}{\xi}
\right]
\tau_z
\Psi_{\rm MF}.
\end{eqnarray}
In this expression
$\Psi_{\rm MF}$
is defined by
Eq.~(\ref{spinor_def}),
and the coefficient
$B_{\rm e}$
is real. If $N$ is not too large, the maximum of the wave function weight
$|\psi_{\rm e}|^2$
is at
\begin{eqnarray}
r \sim r^* = \frac{2l_{\rm B}^2}{\xi} = \frac{2 c \Delta}{ e v_{\rm F} B}.
\end{eqnarray}
This means that Majorana fermion
$\psi_{\rm e}$
localizes along a circle centered at the vortex core, with field-dependent
radius
$r^*(B)$.
We will refer to this state as the exterior MF (thus, the subscript `e').

There is an important distinction between these two MFs. The state at the
vortex core
$\psi_{\rm v}$
is localized by `the vorticity'. That is, as long as the vortex is present,
the wave function
$\psi_{\rm v}$
remains normalizable even for zero magnetic field
($l_{\rm B} \rightarrow \infty$),
with its weight mostly confined within a circle of radius
$\sim \xi$,
the latter quantity being field-independent. Because of this, it is
permissible to neglect
${\bf A}$
in the model Hamiltonian. This is a common approximation used in the
literature dedicated to the core-bound Majorana
fermion~\cite{AL,DasSarma1}.
However, including the magnetic field $B$ into the model is of central
importance to study the exterior MF
$\psi_{\rm e}$,
since the MF weight concentrates mostly at a field-dependent radius
$r^* \sim 1/B$.
If $B$ decreases, the radius
$r^*$
grows. Eventually, at some very weak field, the exterior state recedes to
the outer boundaries of the system.

When the value of
${\bar U}$
violates
condition~(\ref{N_int}),
strictly speaking, the Majorana states disappear. Let us consider a weak
violation
of~(\ref{N_int}):
${\bar U} = \sqrt{2N} + \delta {\bar U}$,
where
$\delta {\bar U}=\delta U/\hbar\omega_c$.
Treating the term
$\delta U \tau_z$
as a perturbation, one can evaluate the matrix element
$\delta U \langle \psi_{\rm v}| \tau_z | \psi_{\rm e} \rangle$
and obtain the corresponding energy splitting
\begin{equation}\label{splitEn}
\delta E \approx \delta U\sqrt{\frac{l_B}{\xi}}\exp{\left(-\frac{l_B^2}{\xi^2}\right)}.
\end{equation}
This value characterizes the hybridization between the core and exterior MF
states for non-zero
$\delta {\bar U}$.
Because of this hybridization, two MFs are replaced by a single Dirac
fermion with energy
$\delta E$.
The eigenenergy is an oscillating function of $U$, vanishing each time when
condition~(\ref{N_int}) is met. In the limit
$\xi \ll r^*$
($\xi \ll l_{\rm B}$),
the overlap between the Majorana wave functions is exponentially small, and
the hybridization may be neglected.

These overlap oscillations are not uncommon in the Majorana fermion
physics: similar phenomena were discussed in other systems as
well~\cite{maj_tunneling,zvyagin}.
It is interesting that single-parameters tuning is sufficient to nullify the
hybridization between the MFs. This is unlike a common two-level system,
which avoids level crossing unless multiparameter fine-tuning is performed.
Such a deviation from a generic behavior occurs because a single (real)
parameter
$t_{12}$
is enough for the complete specification of the most general Hamiltonian
\begin{eqnarray}
H_{12} = i t_{12} \hat{\gamma_1} \hat{\gamma_2},
\end{eqnarray}
describing two coupled Majorana fermions
$\hat\gamma_{1,2}$.
Thus, single-parameter fine-tuning condition
$t_{12} = 0$
is sufficient to guarantee the nullification of the Hamiltonian, and the
resultant level crossing at zero eigenenergy.

\section{Spinless two-dimensional $P$-wave superconductor}
\label{sec::pwave}

Two-dimensional $p$-wave
superconductor~\cite{SrRuO}
Sr$_2$RuO$_4$
is another system where an exterior MF can emerge. The relevant description
is very similar to the calculations presented above. We write down the
model's Hamiltonian in the
form~\cite{Alicea,maj_tunneling}
\begin{equation}
\label{Hk}
H=\left(
          \begin{array}{cc}
            \frac{(i\hbar\nabla-e\mathbf{A}/c)^2}{2m}-U & -\frac{i\hbar}{p_F}
		\{\Delta,\partial_z\}\\
            \frac{i\hbar}{p_F}\{\Delta^*,\partial_z^*\} & -\frac{(i\hbar\nabla+e\mathbf{A}/c)^2}{2m}+U \\
          \end{array}
        \right).
\end{equation}
Here
$m$
is the electron mass,
$p_F=mv_F$
is the Fermi momentum, the anticommutator is defined in the usual manner:
$\{\Delta,\partial_z\} = \Delta \partial_z + \partial_z \Delta$,
where
\begin{eqnarray}
\partial_z=e^{i\phi}\left(\partial_r+ir^{-1}\partial_\phi\right).
\end{eqnarray}
The order parameter, as before, is given by
Eq.~(\ref{order_parameter}).
For the case of a single vortex
($l=1$)
we seek zero-energy eigenfunctions in the form
\begin{eqnarray}
\psi_{\rm MF}
=
\left(
	e^{i(\phi-\pi/4)}u_{\textrm{MF}},e^{-i(\phi-\pi/4)}v_{\textrm{MF}}
\right)^T.
\end{eqnarray}
Similar to previous consideration, we derive a system of differential equations for the radial part of
$\psi_{\rm MF}$.

Introducing the particle-hole conjugation operator for the
$p$-wave superconductor as
$\Xi = \tau_x K$,
one can prove that Hamiltonian in
Eq.~\eqref{Hk}
possesses particle-hole symmetry:
$\Xi H \Xi = - H$.
This property, together with the fact that the differential equations for
the radial part of the wave function are real, implies that the MF radial
wave function satisfies the conditions
$u_{\rm MF} = \lambda v_{\rm MF}$,
where
$\lambda = \pm 1$.
Using these relations we derive the differential equation for function
$\chi$
\begin{eqnarray}
\label{chi_eq_pwave}
\chi''+\frac{\chi'}{\bar r}
+
\chi\!\!
\left({\bar U}\!-\!\alpha^2f^2\!-\!\frac{1}{\bar r^2}\!
-\!\frac{\bar r^2}{4}\!-\!\frac{1}{2}\right)
=0,
\end{eqnarray}
which is connected to the MF wave function as follows
\begin{eqnarray}
\label{chi_eq_pwave1}
u_{\rm MF} = \chi({\bar r}) \exp	\left[
					\lambda\alpha
					\int_0^{\bar r}
						f({\bar r}')d{\bar r}'
				\right],
\\
\text{where}
\quad
{\bar r}=\frac{r}{l_{\rm B}},
\quad
{\bar U}=\frac{2ml_{\rm B}^2U}{\hbar^2}
=
\frac{2 U}{\hbar \omega_{\rm c}},
\end{eqnarray}
and the cyclotron frequency for a massive particle equals to
$\omega_{\rm c} = eB/mc$.

Approximating
$f(r) \approx 1$
we notice that
Eqs.~(\ref{chi_eq_pwave})
and~(\ref{chi_eq}) have the same structure. Exploiting this, one can prove
that the
Hamiltonian~(\ref{Hk})
admits two MF solutions if
\begin{eqnarray}
\label{no_splitting}
{\bar U}-\alpha^2-\frac{1}{2}
=
\frac{2U}{\hbar \omega_{\rm c}}
-
\frac{l_{\rm B}^2}{\xi^2}
-
\frac{1}{2}
= 2N,
\end{eqnarray}
where
$N$
is non-negative integer. From
Eq.~\eqref{chi_eq_pwave1}
it is easy to see that the core-localized MF corresponds
$\lambda=-1$,
while the exterior MF corresponds
$\lambda = 1$.
Up to a normalization coefficient, the MF wave functions are
\begin{eqnarray}
\label{majorana_v}
\psi_{\rm v,e}\!=\!
\frac{r}{2l_{\rm B}^2}
\exp{\!\!\left(\!\frac{\lambda \Delta}{v_F}\!\int\limits_0^r\!\! fdr'\!-\!\frac{r^2}{4l_{\rm B}^2}\right)}
L_N^{(-1)}\!
 \begin{pmatrix}
    e^{i(\phi-\frac{\pi}{4})}
    \\
   \lambda e^{-i(\phi-\frac{\pi}{4})}
  \end{pmatrix}.
\quad
\end{eqnarray}
One can verify that
$\Xi \psi_{\rm v,e}= \lambda \psi_{\rm v,e}$,
and both MF wave functions are orthogonal to each other. Hence,
$\psi_{\rm v,e}$
are two independent MFs. When the
condition~(\ref{no_splitting})
is violated, the MFs are hybridized, forming a single Dirac electron as in
the previously discussed case of the TI/SC heterostructure.

We assumed above that
$f(r) \approx 1$,
while the function $f$ deviates from unity near the core. This deviation
may be accounted using perturbation theory. As a result, the
condition~(\ref{no_splitting})
and the wave
functions~(\ref{majorana_v})
will be slightly corrected.

\section{Discussion}
\label{sec::disco}

It is typically assumed that, in order to localize a MF, one needs a
boundary separating two parts of the system with different topological
numbers or some ``topological" defect. Two models considered above serve as
counterexamples to this statement: we have shown that the exterior Majorana
wave function is not ``latched" to any inhomogeneity. Instead, the
localization radius
$r^*$
is a magnetic field-dependent quantity, and may be manipulated in real time.

The latter feature allows us to control the coupling between the MFs. Let
two vortices are pinned at distance $R$ from each other. Each vortex hosts
a core-localized state and an exterior Majorana states. Straightforward
calculations show that the splitting between the exterior MF of the first
vortex and the core MF of the second vortex is zero. However, the splitting
between the exterior MFs of different vortices is non-zero and depends on
$B$. When the magnetic field is high, such that
$R \gg r^* = 2l_{\rm B}^2/\xi \sim 1/B$,
the coupling between two exterior MFs is exponentially weak, and can be
neglected. With the decrease of $B$ the localization radius
$r^*$
grows, and so does the coupling between the MFs.

Changing the hybridization between the MFs allows one to perform braiding
without moving the vortices. While the topological computations are not the
main topic of this letter, we briefly outline a possible braiding protocol
similar to one proposed in
Ref.~\onlinecite{Clarke_braid}.
Consider first the motion of a single MF. Initially
($t=0$)
we have one separate MF
$\gamma_1$
and one Dirac fermion, the latter consisting of two coupled MFs
$\gamma_2$
and
$\gamma_3$.
Transportation of the MF
$\gamma_1$
from its initial position to the position of
$\gamma_3$
can be described by the Hamiltonian
\begin{eqnarray}
\label{tun}
H(t)
=
\zeta_{12}\alpha(t)\gamma_1\gamma_2
+
\zeta_{23}[1-\alpha(t)]\gamma_2\gamma_3,
\end{eqnarray}
where
$\zeta_{ij}$
are the tunneling amplitudes between the $i$-th and $j$-th MFs. The
coefficient
$\alpha (t)$
changes adiabatically from
$\alpha(0)=0$
to
$\alpha(t_1)=1$.
The equation of motion
\begin{eqnarray}
\dot{\gamma_i}=i[H(t),\gamma_i(t)]
\end{eqnarray}
can be written as
\begin{eqnarray}
\dot{\gamma_a}=2\epsilon_{abc}B_b\gamma_c,
\end{eqnarray}
where the vector
${\bf B}$
is equal to
\begin{eqnarray}
{\bf B}(t) =
[1-\alpha(t)]\zeta_{23}(1,0,0)+\alpha(t)\zeta_{12}(0,0,1),
\end{eqnarray}
and
$\epsilon_{abc}$
is the antisymmetric Levi-Civita tensor. Straightforward calculations give
the solution
\begin{eqnarray}
\label{eq::transport}
\gamma_3(t_1)=\textrm{sgn}(\zeta_{12}\zeta_{23})\gamma_1(0).
\end{eqnarray}
Thus, the MF can be transported by tuning the interaction between MFs. For
exterior MFs, this interaction may be varied by adjusting local magnetic
field.

\begin{figure}[t!]
\center
\includegraphics[width=7 cm, height=5 cm]{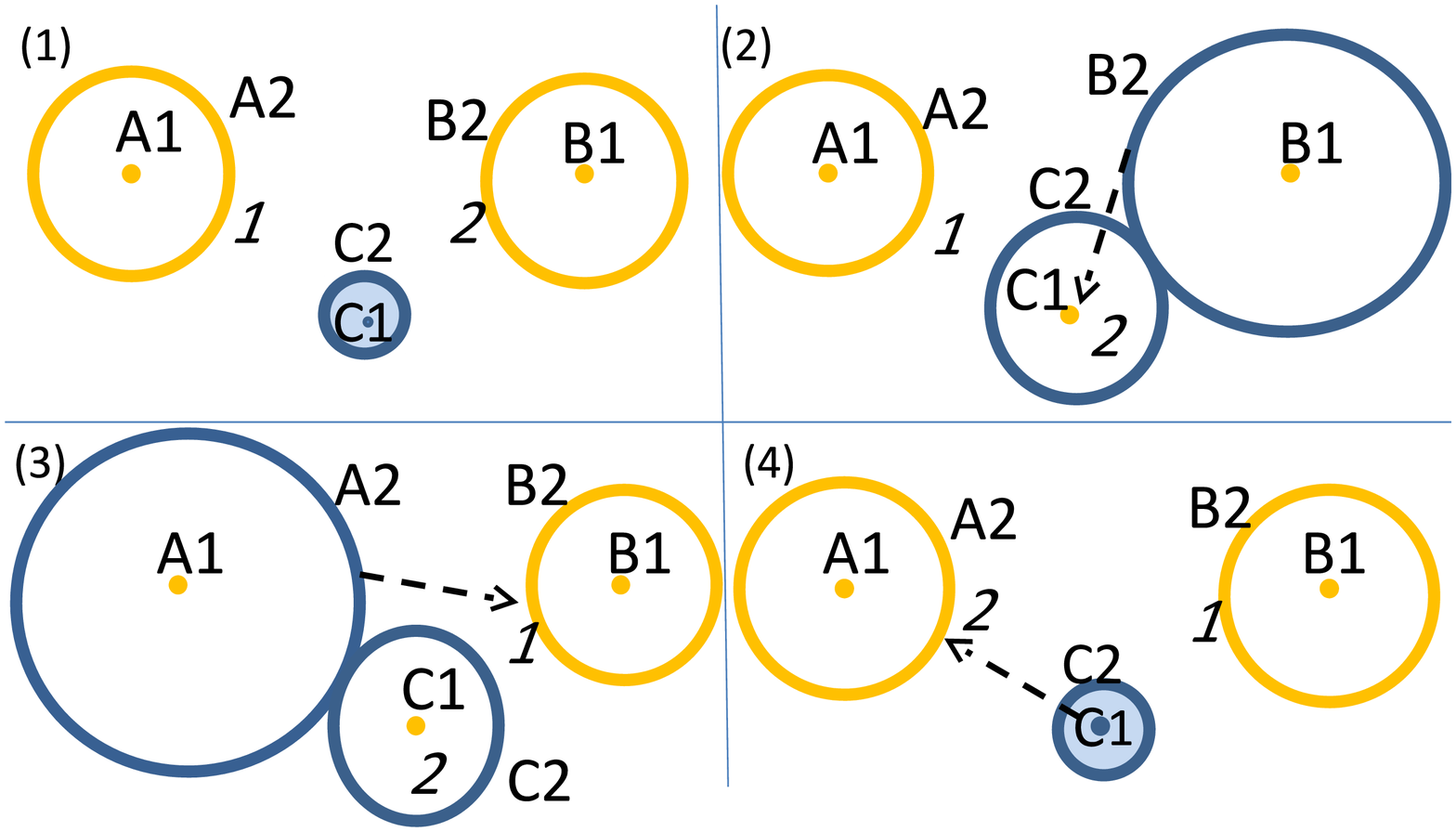}
\caption{(Color online)
Braiding of two MFs. The consecutive steps are shown in panels (1-4). Each
step or shift is a transportation of a single MF between two positions [see
discussion for
Eqs.~(\ref{tun})
and~(\ref{eq::transport})].
This move (or shift) is represented by a dashed arrow. Orange circles
(points) correspond to an exterior (core) MF. When two MFs couple and form a
single Dirac fermion, the color is changed to blue. Initially [panel (1)]
we have two exterior MFs (1 and 2) on vortices
$A$
and
$B$.
The MFs of
vortex
$C$
are coupled into a single Dirac state. In panel (4) everything
is the same except that MFs 1 and 2 switched their positions.
\label{braid}
}
\end{figure}

The procedure described above can be used to implement braiding of two
exterior MFs. As a result of the braiding protocol execution, two exterior
Majorana fermions swap their positions. The protocol requires three pinned
vortices. One vortex is auxiliary, two other serve as the starting and
ending locations for the Majorana fermions (denoted below as MF~1 and MF~2)
participating in the braiding, see
Fig.~\ref{braid}.
Initially, MF~1 (MF~2) is in position
$A2$
($B2$),
while the MFs associated with vortex $C$ are coupled to form a single Dirac
fermion. Next, MF~$2$ is transported to
$C1$.
After that MF~1 goes to
$B2$.
Finally, MF~$2$
moves to
$A2$,
while MF
$1$
remains in
$B2$.

In our consideration of the TI/SC heterostructure, we assumed that the magnetic field is uniform, while the suppression of the superconductivity by this field is negligible. Therefore, the applied field must be much smaller than the second critical field of the superconducting layer, that is,
$l_{\rm B} \gg \xi_{\rm SC}$.
Under this condition, the regime
$l_{\rm B} \sim \xi$
can be achieved if
$\xi \gg \xi_{\rm SC}$.
Taking the characteristic values
$|\Delta| = 1$~meV and
$v_\textrm{F}=5 \times 10^7$~cm/s,
we obtain
$\xi \approx 400$~nm.
The condition
$l_{\rm B} \sim \xi$
corresponds to
$B \sim 4$~mT,
which is much smaller than the second critical magnetic field for Pb or Nb.
Even in the case of smaller coherence length,
$\xi \approx 100$~nm,
we have
$B \sim 64$~mT,
which is still smaller than the critical magnetic field for Pb or Nb. For
the latter case, we can estimate the value of the Zeeman splitting as
$V_z=g\mu_{\rm B}B/2 \sim 0.1$~meV
(where
$g=50$
is the Land\'{e} factor and
$\mu_B$
is the Bohr magneton). This energy scale is much smaller than the
superconducting order parameter and can be neglected, as we assumed above.
Finally, in thin films the London penetration depth is large
($\lambda_{\rm L} \sim 12,000$~nm, see
Ref.~\onlinecite{Rodichev}).
Consequently, the assumption that $B$ is uniform is reasonable.

\section{Summary}
\label{sec::summary}

We analyzed the generation of MFs in topological superconductors in the
presence of vortices. Discussing both (1) the topological insulator -
superconductor heterostructure and (2) a spinless $p$-wave superconductor,
we established that the combined effect of the vortex and the transverse
magnetic field $B$ gives rise to the generation of two MFs. The first MF is
well-known: it localizes near the vortex core. The second, exterior, MF is
localized by the magnetic field.  Its wave function weight is centered
along a circle of radius
$r^*\propto 1/B$.
Varying the magnetic field we can change the positions of the MFs and tune
the splitting between the core and exterior MFs of the same vortex or
between exterior MFs of different vortices in real time. The tunability of
the pairwise couplings between MFs by a magnetic field may open a new route
for future topological quantum operations.

\vspace{0.5 cm}
\section*{Acknowledgements}
We acknowledge partial support from
the Dynasty Foundation and ICFPM (MMK), the Ministry of Education and
Science of the Russian Federation Grant No. 14Y26.31.0007, RFBR Grant No.
15-02-02128.
F.N. was partially supported by: the RIKEN iTHES Project, the MURI Center
for Dynamic Magneto-Optics via the AFOSR Award No. FA9550-14-1-0040, the
Japan Society for the Promotion of Science (KAKENHI), and a grant from the
John Templeton Foundation.

\end{document}